\documentclass{article}
\usepackage{amsmath}
\usepackage{amsfonts}
\usepackage{amssymb}
\newtheorem{theorem}{Theorem}[section]

\newtheorem{lemma}[theorem]{Lemma}

\newtheorem{corollary}[theorem]{Corollary}

\begin{document}
\title{On eigenvalues of discrete Schr\"odinger
operators with potentials of Coulomb type decay
}
\author{Denis Krutikov}
\maketitle

\noindent Universit\"at Essen, Fachbereich Mathematik/Informatik,
45117 Essen,

\noindent GERMANY\\
E-mail: denis.krutikov@uni-essen.de\\[0.3cm]
\\[0.3cm]
\\[0.3cm]
\begin{abstract}
We study the distribution of the eigenvalues inside of the essential spectrum for
discrete one-dimensional Schr\"odinger operators with potentials
of Coulomb type decay.

\end{abstract}
\section{Introduction}
In this paper we study one-dimensional discrete Schr\"odinger
operators on the "half line" (that is on $ \ell_2(\mathbb{N})$),
which are defined by
\begin{equation}
(H_{\varphi}y) ( n)=y(n-1)+y(n+1)+V(n)y(n)
\nonumber
\end{equation}
(where $0 < \varphi < \pi$) along with a phase boundary condition
\begin{equation}
y(0) \sin \varphi+ y(1) \cos \varphi=0.
\nonumber
\end{equation}
The actual value $\varphi$ from the definition
of the operator $H_{\varphi}$ will not be significant,
therefore we will omit the index
$\varphi$ and will write $H$ instead of $H_{\varphi}$.\\
We assume that $V(n)$ is a Coulomb potential, that is $V(n)$ satisfies
\begin{equation}
\label{con} |V(n)| \le \frac{C}{n} \end{equation}
for some
constant $C>0$. We prove below the following theorem:
\begin{theorem}
\label{t1} Let a potential $V(n)$ satisfy the condition
(\ref{con}).  Let $E_j$,\\ $j=1,2,...$, be eigenvalues of
$H$ corresponding to $l^2$-eigenvectors. Suppose
all $E_j$'s lie in $(-2,
2)$ and $E_j \neq E_k$ if $j \neq k$. Then holds the following
inequality: \begin{equation} \label{eqt.1} \sum_{j=1}^{\infty}
\left(1-\frac{E^2_j}{4} \right) \le \frac{C^2+2}{2}.
\end{equation}
In particular, the eigenvalues which lie inside of the interval
$[-2,2]$ (which is the essential spectrum of $H$ in the case under
consideration, see \cite{Te}, Chapter 3) form a finite or a countable set
with only two possible accumulation points $-2$ and $2$.
\end{theorem}

\vspace{1mm}

\noindent
{\it Remark.}
The similar result for continuous one-dimensional Schr\"odinger
operators with potentials of Coulomb type decay was proved
by Kiselev, Last and Simon in \cite{KLS}, and the similar result for Dirac operators
with Coulomb potentials was proved by the author in \cite{Kr}.

\section{Auxiliary results}
We use a EFGP transformation (also called a Pr\"ufer
transformation) to rewrite the discrete Schr\"odinger equation
\begin{equation}
\label{scheq} y(n-1)+y(n+1)+V(n)y(n)=Ey(n)\quad\quad\quad
(n\in\mathbb N)
\end{equation}
for $E$ from the interval $(-2,2)$.
So, suppose that $E \in (-2,2)$ and let $y$ be some solution of
(\ref{scheq}). Write $E=2\cos x$ with $x \in (0,\pi)$ and define
$R(n)>0$, $\theta(n)$ by
\[
\begin{pmatrix}
u(n)-u(n-1) \cos x \\ u(n-1) \sin x \end{pmatrix}
= R(n) \begin{pmatrix} \cos(\theta(n)) \\
 \sin(\theta(n)) \end{pmatrix} .
\]
(We note that EFGP variables $R(n)$ and $\theta(n)$ depend on the
spectral parameter $x$.) Denote $\frac{V(n)}{\sin x}$ with
$\nu_x(n)$. Then $R$ and $\theta$ obey the equations (see
\cite{KLS})
\begin{equation}
\label{EFGP0} R(n)^2= u(n)^2+u(n-1)^2-2 u(n) u(n-1) \cos x ,
\end{equation}
\begin{equation}
\label{EFGP1}
\frac{R(n+1)^2}{R(n)^2}=
\left(1-\nu_x(n) \sin(2\theta(n)+2x)+ \nu^2_x(n) \sin^2(\theta(n)+x) \right),
\end{equation}
\begin{equation}
\label{EFGP2}
\cot (\theta(n+1))=\cot(\theta(n)+x)-\nu_x(n).
\end{equation}

\begin{lemma} (due to \cite{KLS}, \cite{KRS})
\label{l2.1}
\hspace{1mm} If \, $|\nu_x(n)| < \frac{1}{2} $, then
\begin{equation}
|\theta(n+1) - \theta(n) - x| \le \pi |\nu_k(n) |.
\end{equation}

\end{lemma}

\begin{lemma}
\label{l2.2}
Let $(\gamma_n)$ be a sequence of real numbers with the property \\
$|\gamma_{n+1}-\gamma_n| \le C_1/n$ for some constant $C_1>0$
for $n> n_0$ and let $\alpha $
be some real number $\neq 2 \pi n$, $n \in \mathbb{Z}$. Then the
sequence
 \[ \left( \sum_{n=1}^N \frac{1}{n} e^{i (\alpha n +\gamma_n)}
 \right)_{N=1}^{\infty} \] is bounded.
\end{lemma}

\noindent {\it Proof.} We can assume without loss of generality
$n_0=1$.
We use the Abel transformation to obtain
\[\sum_{n=1}^N \frac{1}{n} e^{i (\alpha n +\gamma_n)} =
\frac{1}{N} e^{i \gamma_N} \sum_{j=1}^N
e^{i \alpha n}- \sum_{n=1}^{N-1}
\left( \frac{1}{n+1} e^{i \gamma_{n+1}}
-  \frac{1}{n} e^{i \gamma_n}
\right)  \sum_{j=1}^n e^{i \alpha j}.
\]
The first summand is bounded as $N \rightarrow \infty$ because of \[\sum_{j=1}^n e^{i \alpha
j} =  e^{i \alpha } (1- e^{i \alpha n}) (1- e^{i \alpha})^{-1}\]
and the second one is bounded as $N \rightarrow \infty$ because the series \\
\[ \sum_{n=1}^{\infty} \left( \frac{1}{n+1} e^{i \gamma_{n+1}} -
\frac{1}{n} e^{i \gamma_n} \right)  \sum_{j=1}^n e^{i \alpha j} \]
converges, for we can majorize $| \frac{1}{n+1} e^{i \gamma_{n+1}}
- \frac{1}{n} e^{i \gamma_n} | $ by $(1+C_1)/n^2 $ because of
\begin{multline*} \left| \frac{1}{n+1} e^{i \gamma_{n+1}}
-  \frac{1}{n} e^{i \gamma_n} \right|= \left|\frac{1}{n(n+1)} e^{i
\gamma_n} (ne^{i (\gamma_{n+1}-\gamma_n)} -n-1)\right|
\\ \le \frac{1}{n^2} \left( n |e^{i (\gamma_{n+1}-\gamma_n)}
-1|+1 \right) \le \frac{1}{n^2}+\frac{1}{n} \sup\limits _{ |x| \le
C_1/n } |e^{i x} -1| \\= \frac{1}{n^2}+ \frac{1}{n} |e^{i C_1/n}
-1| = \frac{1}{n^2}+ \frac{1}{n} 2 |\sin \frac{C_1}{2n}| \le
\frac{C_1+1}{n^2}. \hspace{5mm} \square
\end{multline*}
\begin{corollary}
\label{c2.3} Let $x_j$, $j=1,...,m$, be real numbers which satisfy
$2 x_j \neq \pi n $, $n \in \mathbb{Z}$, for all $j$ and $x_j \pm
x_k \neq \pi n $, $n \in \mathbb{Z}$, for all $j \neq k$, and let
$\theta_j(n)$, $j=1,...,m$, be EFGP angles corresponding to the
eigenvalues $2 \cos x_j$ of $H$. Denote $\theta_j(n)+x_j$ with
$\bar \theta_j(n)$. Then the sequences
\begin{equation}
\label{c1} \left( \sum_{n=1}^N \frac{1}{n} \sin 2 \bar \theta_j(n)
\sin 2 \bar \theta_k(n) \right)_{N=1}^{\infty}, \hspace{3mm} j
\neq k,
\end{equation}
and
\begin{equation}
\label{c2}
 \left( \frac{\ln N}{2} - \sum_{n=1}^N \frac{1}{n}
\sin 2 \bar \theta_j^2(n) \right)_{N=1}^{\infty}
\end{equation}
are bounded.
\end{corollary}

\noindent {\it Proof.} First of all we choose $n_0$ so that holds
$|\nu_{x_j}(n)| < 1/2$ for all $j=1,...,m$ and for all $n>n_0$. We
can do it because of the condition (\ref{con}) (note that $\min_{j=1,...,m} |\sin x_j|
>0$).\\
Then we use Lemma \ref{l2.2} with three different definitions of $\gamma_n$ and $\alpha$: \\
1)$\gamma_n:=2(\bar \theta_j(n)+\bar \theta_k(n)) -2x_j n - 2x_k n$ and $\alpha:=2x_j+2x_k$ ($j \neq k$),\\
2)$\gamma_n:=2(\bar \theta_j(n)-\bar \theta_k(n)) -2x_j n + 2x_k n$ and $\alpha:=2x_j-2x_k$  ($j \neq k$) and \\
3)$\gamma_n:=4\bar \theta_j(n) -4x_j n$ and $\alpha:=4x_j$. \\
In all three cases the condition $|\gamma_{n+1}-\gamma_n| \le
\frac{4 \pi C}{a n}$ is satisfied for $n>n_0$, because of Lemma \ref{l2.1} and
the condition (\ref{con}). (We note also that $\alpha$'s satisfy in
all three cases the condition $\alpha \neq 2 \pi n$.)
\par We consider only the
real parts of $\left( \sum_{n=1}^N \frac{1}{n} e^{i (\alpha n
+\gamma_n)} \right)$ to
obtain that the following three sequences are bounded:\\
$\sum_{n=1}^N \frac{1}{n}
\cos (2 \bar \theta_j(n) + 2 \bar \theta_k(n))$ (the case 1),\\
$\sum_{n=1}^N \frac{1}{n}
\cos (2 \bar \theta_j(n) - 2 \bar \theta_k(n))$ (the case 2) and\\
 $\sum_{n=1}^N \frac{1}{n} \cos (4 \bar \theta_j(n)
)=\sum_{n=1}^N \frac{1}{n} (1-2\sin^2 2 \bar \theta_j(n) ) $ (the
case 3). \par Then we obtain the boundedness of (\ref{c1}) by
substraction of the first sequence  from the second one and the
boundedness of (\ref{c2}) from the boundedness of the third one by
the boundedness of the sequence $\left( \sum_{n=1}^N \frac{1}{n} -
\ln N \right) $. \hspace{5mm} $\square$

\begin{lemma}(due to \cite{KLS})
\label{l2.4} Let $\{e_i\}^N_{i=1}$ be such a set of unit vectors in a
Hilbert space $\cal{H}$, that holds
\[ \beta: \, =\sup\limits_{k \neq j} \langle e_k,e_j \rangle _{\cal{H}} < 1/N. \]
Then for any g from $\cal{H}$ holds
\[ \sum^N_{j=1}|\langle g, \, e_j \rangle _{\cal{H}}|^2 \le (1+\beta N)\|g \|_{\cal{H}}^2.\]
\end{lemma}

\begin{lemma}
\label{l.ln}
For all $x$ from $(0,\varepsilon)$ ($\varepsilon>0$) hold the inequalities
\[ \ln(1+x) \ge \frac{1}{1+\varepsilon} x
\]
and
\[ \ln(1-x) \ge -\frac{1}{1-\varepsilon}x.
\]
\end{lemma}
{\it Proof.} We have only to use Mean Value Theorem: \[
\ln(1+x)=\ln(1+x) - \ln 1 \ge x \inf\limits_{\xi \in (1, 1+x)}
\frac{1}{\xi} \ge x \frac{1}{1+\varepsilon},\hspace{4mm} \]
\[  -\ln(1-x)=\ln(1) - \ln (1-x) \le x
\sup\limits_{\xi \in (1-x, 1)} \frac{1}{\xi}  \le x
\frac{1}{1-\varepsilon}. \hspace{4mm} \square\]

\section{Proof of Theorem \ref{t1}}
Let $m$ be an arbitrary but fixed positive integer. We consider
eigenvalues $E_1,...,E_m$. Write $E_j=2 \cos x_j$ with $x_j \in
(0, \pi)$.
Denote $\min_{j=1,...,m} \sin x_j$ with $a$. It is evident that $a>0$.\\
We assume for the moment that all $x_j$'s lie in the interval $(0,
\pi/2)$.

\vspace{1mm}

\noindent Let EFGP-variables $R_j(n)$ and $\bar \theta_j(n)$
correspond to the $l^2$-solution $y(n)$ of (\ref{scheq}) with
$E_j=2 \cos x_j$ (we assume that $R_j$ is normalized by
$R_j(0)=1$). From (\ref{EFGP0}) follows $ \sum_{j=1}^m R_j(n)^2
\in l^1 $, which implies
\[ \underline \lim \, \, n \sum_{j=1}^m R_j(n)^2 =0.
\]
Thus, there exists the sequence $N_l \rightarrow \infty$
(monotonically) so that for $j=1,...,m$ and for all $l$ holds
\begin{equation}
\label{eq}
R_j(N_l)^2 \le \frac{1}{N_l}.
\end{equation}
Now we choose $n_0$ so that for $n>n_0$ holds the inequality $C/an < 1/2$, which implies
$| \nu_{x_j}(n)| < 1/2$ for all $j=1,..,m$.
Using (\ref{EFGP1}) we obtain from (\ref{eq}):
\[ \ln (R_j(n_0)) + \sum_{n=n_0}^{N_l-1} \ln (1- \nu_{x_j}(n) \sin (2 \bar \theta_j(n))+
\nu_{x_j}(n)^2 \sin^2 (\bar \theta_j(n))) \le - \ln N_l.\] Using
Lemma \ref{l.ln} and Lemma \ref{l2.1}, the condition (\ref{con}) and the monotone behavior of the function $\ln x
$, we obtain from the last inequality for every $j$:
\begin{equation}
\label{eq2} \ln (R_j(n_0))  -  \sum_{n=n_0}^{N_l-1} a_{j, \, n}
\nu_{x_j}(n) \sin (2 \bar \theta_j(n)) \le - \ln N_l,
\end{equation}
where $a_{j,n}$ is defined by

\[ a_{j, \, n}  = \left\{ \begin{array}{c}
\frac{1}{1-C/a n} \, , \hspace{4mm} \nu_{x_j}(n)  \sin (2 \bar \theta_j(n)) \ge 0
\\

\vspace{1mm}

\frac{1}{1+C/a n} \, ,
 \hspace{4mm}  \nu_{x_j}(n)  \sin (2 \bar \theta_j(n)) <0
. \end{array} \right\}
\]
From the condition (\ref{con}) follows easily the convergence of
the series
\[ \sum_{n=n_0}^{\infty}
(a_{j, \, n}-1) V(n) \sin (2 \bar \theta_j(n)),
\]
which implies (taking in account (\ref{eq2})) that there exists
such a constant $C_2 $, that for all $j$ and all $l$ holds the
following inequality:
\begin{equation}
\label{ineq1} \sin x_j \ln N_l - \sum_{n=n_0}^{N_l-1}
 V(n) \sin (2 \bar \theta_j(n)) \le C_2.
\end{equation}
Now we define for each fixed $l$ the Hilbert space $\cal{H}$$_l$
as follows: \\$\cal{H}$$_l$: $=$ $l^2(n_0,...,N_l-1)(n)$, that is
the set of all finite sequences $(b_n)_{n=n_0}^{N_l-1}$ with the
scalar product $ \langle b, \, c \rangle = \sum_{n=n_0}^{N_l-1} n
b_n c_n$ (where $b=(b_n)$, $c=(c_n)$).

\noindent We define vectors $f_j(n)$, $e_j(n)$ ($\in \cal{H}$$_l$)
by $f_j(n) \, =\frac{\sin 2 \bar \theta_j(n)}{n}$, $e_j(n) \, =
\frac{f_j(n)}{\| f_j(n)\|}$. From $x_j \neq x_k$, $j \neq k$, and
$x_j \in (0, \pi/2)$ follows $4x_j \neq 2 \pi n$, $2x_j \pm 2 x_k
\neq 2 \pi n $, $j \neq k$, so we can apply Corollary \ref{c2.3}
to obtain from (\ref{c2})
\[\| f_j(n)\|^2 = \sum_{n=n_0}^{N_l-1} n^{-1} \sin^2 2 \bar
\theta_j(n) = \ln N_l /2 +O(1).\]
($O(1)$ denotes here and further any quantity
which is bounded as $l \rightarrow \infty$.)
\par If $j \neq k$ we have from
(\ref{c1}):
\[\langle e_j, \, e_k \rangle = \| f_j(n)\|^{-1} \| f_k(n)\|^{-1}
\sum_{n=n_0}^{N_l-1} n^{-1} \sin 2 \bar \theta_j(n) \sin 2 \bar
\theta_k(n) =  \frac{1}{ \ln N_l} O(1). \] So we can use Lemma
\ref{l2.4} with $g: \, =V(n) \in \cal{H}$$_l$ with sufficiently
large $N_l$ (so that for $\frac{1}{ \ln N_l} O(1)$ from the last
formula holds $\frac{1}{ \ln N_l} O(1) <1/m$) to obtain the
inequality
\[ \sum_{j=1}^m \left( \sum_{n=n_0}^{N_l-1}
n V(n) e_j(n) \right)^2 \le \left( 1 + m \frac{1}{ \ln N_l} O(1)
\right) \sum_{n=n_0}^{N_l-1} n V(n)^2.
\]
From the last inequality together with (\ref{ineq1}) and with the
inequality \[ \sum_{n=n_0}^{N_l-1} n V(n)^2  \le C^2 (\ln N_l +
O(1))\] (where we used the condition (\ref{con})) follows now for
sufficiently large $N_l$'s:
\begin{multline*} \sum_{j=1}^m \left(\sin x_j \ln N_l -C_2\right)^2 \le
\sum_{j=1}^m  \left( \sum_{n=n_0}^{N_l-1}
 V(n) \sin (2 \bar \theta_j(n))\right)^2 \\= \sum_{j=1}^m
\left(\sum_{n=n_0}^{N_l-1}
 n V(n) f_j(n)\right)^2
= \sum_{j=1}^m  \| f_j(n)\|^2 \left(\sum_{n=n_0}^{N_l-1}
 n V(n) e_j(n)\right)^2\\ = \left( \frac{\ln N_l}{2} + O(1)
 \right)
\sum_{j=1}^m \left(\sum_{n=n_0}^{N_l-1}
 n V(n) e_j(n)\right)^2\\
 \le
 \left( \frac{\ln N_l}{2} + O(1)
 \right)
 \left( 1 + \frac{m }{ \ln N_l}O(1)
\right) C^2 (\ln N_l + O(1))\\=\frac{C^2 (\ln N_l)^2}{2}
 \left( 1 + \frac{m}{ \ln N_l} O(1)
 \right).
 \end{multline*}
 On the other hand we have
\[\sum_{j=1}^m \left(\sin x_j \ln N_l -C_2\right)^2 = (\ln N_l)^2
\left(\frac{m}{\ln N_l} O(1)+ \sum_{j=1}^m \sin^2 x_j \right).
\]
As we can choose $N_l$ arbitrarily large, we obtain then $
\sum_{j=1}^m \sin^2 x_j \le \frac{C^2 }{2}$.

\vspace{1mm}

\par It is easy to see
that if we replace the assumption $x_j \in (0, \pi/2)$ for all \\
$j=1,...,m$ by the new assumption $x_j \in (\pi/2, \pi)$ for all
$j=1,...,m$, the whole consideration remains valid (we have in
this case also $2x_j \pm 2 x_k \neq 2 \pi n $, $j \neq k$, and
$4x_j \neq 2 \pi n$). Returning to a general situation we have now
to take in account the ''critical'' point $\pi/2$. Because of
$\sin (\pi/2)=1$ we obtain in a general situation the inequality
\[ \sum_{j=1}^m \sin^2 x_j \le \frac{C^2 }{2}+1. \]
The right side of this inequality is independent of $m$, therefore
we can replace the finite sum on the left with the infinite sum
$\sum_{j=1}^{\infty}
\sin^2 x_j$. We have now only to use the relation
$\sin^2 x_j=1-\cos^2 x_j=1-E_j^2/4$.
The inequality (\ref{eqt.1}) is proved.

\par From (\ref{eqt.1}) it is easy to see, that for each fixed
$\varepsilon$ from $(0, 1)$ there exists at most finite number of
eigenvalues $E_j$ with $E_j \in (-2+\varepsilon, 2- \varepsilon)$.
So the only possible accumulation points of the set of eigenvalues
of $H$ lying in the interval $[-2,2]$ are the points $2$ and $-2$.
\hspace{5mm} $\square$

\end{document}